\documentclass[review]{elsarticle}
\usepackage{graphicx}
\usepackage{amssymb}
\usepackage{lineno}
\usepackage{lineno,hyperref}
\usepackage{amsmath}
\modulolinenumbers[5]
\usepackage{subfig}
\journal{Ocean Engineering}

\begin{document}
\begin{frontmatter}


\title{The effects of free stream turbulence on the hydrodynamic characteristics of an AUV hull form}



\author{A. Mitra \fnref{myfootnote}} \author{J.P. Panda} \author{H.V. Warrior}
\address{Department of Ocean Engineering and Naval Architecture,
IIT Kharagpur, India}
\fntext[myfootnote]{corresponding author: arindam.mitra@iitkgp.ac.in}

\begin{abstract}
This article presents experimental and numerical studies on the effect of free stream turbulence on evolution of flow over an autonomous underwater vehicle (AUV) hull form at three Reynolds numbers with different submergence depths and angles of attack. The experiments were conducted in a recirculating water tank and the instantaneous velocity profiles were recorded along the AUV using Acoustic Doppler Velocimetry (ADV). The experimental results of stream-wise mean velocity, turbulent kinetic energy(TKE) and Reynolds stresses were used to validate the predictive capability of a Reynolds stress model (RSM) with the wall reflection term of the pressure strain correlation. From the high fidelity RSM based simulations it is observed that in presence of free stream turbulence, the pressure, skin friction, drag and lift coefficients decrease on the AUV hull. The variation of the hydrodynamic coefficients were also plotted along the AUV hull for different values of submergence depth and angle of attack with different levels of free stream turbulence. The conclusions from this experimental and numerical investigation give guidance for improved design paradigms for the design of AUVs.
\end{abstract}


\end{frontmatter}


\section{Introduction}
\label{S:1}
Mapping and monitoring the marine environment is critically important for a variety of applications. However in many scenarios, like mapping the sea floor, the presence of humans is expensive and impracticable. This necessitates robotic vehicles that can move through the ocean without real-time control by human operators. Such autonomous underwater vehicles (AUVs) are used extensively for underwater
research, environmental monitoring, sea-probes and also for inspection and maintenance of
offshore structures and do not require any direct human control while collecting data \cite{wynn2014autonomous}. These have wide range of applications in defense, oil exploration, and, policy sectors such as geohazard assessment associated with oil and gas infrastructure.

Investigation of the hydrodynamic performance of AUV has substantial significance in the design of AUV. Several experimental and numerical studies are available in the literature in which the evolution of hydrodynamic coefficients over AUV hulls were studied to better understand the hydrodynamic forces and moments acting on the AUV under various conditions\cite{jagadeesh2009experimental,saeidinezhad2015experimental,javadi2015experimental,hayati2013study,papadopoulos2015hydrodynamic,jian2016numerical,rattanasiri2015numerical, gafurov2015autonomous,tyagi2006calculation,allotta2018identification,li2018hydrodynamic}. Jagadees et al.\cite{jagadeesh2009experimental} made experimental analysis of hydrodynamic force coefficients at different angle of attacks and Reynolds numbers over the standard hull form. Saeidinezhad et al.\cite{saeidinezhad2015experimental} analyzed the effect of Reynolds numbers on the pitch and drag coefficient of a submersible vehicle model. Javadi et al.\cite{javadi2015experimental} conducted experimental analysis of the effect of bow profiles on the resistance of the AUV in a towing tank at different Froude numbers and have studied the variability of the friction drag with Froude numbers. Alvarez et al.\cite{alvarez2009hull} examined the wave resistance on the AUV operating near the surface. Wu et al.\cite{wu2014hydrodynamic} explored the hydrodynamics of an AUV approaching the dock at different speeds in a cone-shaped dock under the influence of ocean currents. Tyagi and Sen \cite{tyagi2006calculation} predicted the transverse hydrodynamic coefficients over an AUV hull, that are important in maneuverability study of marine vehicles.

The design of AUVs involves Computational Fluid Dynamics (CFD) simulations, relying on turbulence models to account for the effects of turbulent flow. Consequently the predictive fidelity of different turbulence models for AUV flow simulations is critically important. Several researchers have assessed the performance of different turbulence models in analyzing the hydrodynamic performance of AUV. Jagadeesh and Murali\cite{jagadeesh2005application} have studied the hydrodynamic forces on AUV hull forms using various low-Re version of two-equation turbulence models\cite{jagadeesh2005application}, since those can capture the low turbulence levels in the viscous sublayer and the effects of damped turbulence. Sakthivel et al.\cite{sakthivel2011application} used a nonlinear version of the two equation turbulence model \cite{kimura2003non} to capture the flow physics arising from the cross flow interaction with hull at higher angle of attacks. Jagadeesh et al.\cite{jagadeesh2009experimental} validated the model predictions of \cite{abe1994new} against the experimental results of hydrodynamics coefficients over a AUV hull form. Mansoorzadeh and Javanmard \cite{mansoorzadeh2014investigation} have studied the effect of free surface on drag and lift coefficients on AUV at different submergence depths using both $k-\epsilon$ and shear stress transport(SST) model \cite{ansys} and observed that hydrodynamic coefficients were very responsive to the submergence depth and AUV speed. Salari and Rava \cite{salari2017numerical} studied the hydrodynamics over an AUV near the free surface and have analyzed the wave effects on AUV at different depths from sea surface by using both $k-\epsilon$ and 4 equation turbulence/transition model of\cite{menter2006correlation} which can accurately predict laminar turbulent transitions. Leong et al.\cite{leong2015quasi} analyzed the hydrodynamic interaction effects of AUV nearer to a moving submarine and studied the interactive force and moments. More recently \cite{da2017numerical} used Large Eddy Simulations (LES) to analyze the hydrodynamic coefficients along an AUV with fish tail shape and effect of fish tail shape on the resistance and stability.

A majority researchers have used simple two equation turbulence models for analysis of flow along the AUV, the recent emphasis has been shifted to Reynolds stress models\cite{panda2017,panda2018representation,mishrathesis,mishra1,mishra3}. The turbulent flow over AUV bodies often manifests complexities, such as flow separation or significant streamline curvature. The formulation of eddy viscosity based models involves many simplifications and assumptions that limit their accuracy for such complex turbulent flows\cite{iaccarino2017}. For example, in turbulent flows with significant streamline curvature the predictions of eddy-viscosity based turbulence model is deficient\cite{launder1987}. In turbulent flows with flow separation, linear eddy-viscosity based models are unable to capture the effects of flow separation\cite{craft1996}. For the complex turbulent flow over the AUV body, Reynolds stress models have the potential to provide better predictions than simple two equation eddy-viscosity based models at a computational expense significantly lower than large eddy simulations and direct numerical simulations. The critical component of Reynolds stress models are the pressure strain correlation closures that can capture the directional effects of the Reynolds stresses and additional complex interactions in turbulent flows (\cite{johansson1994}) near the wall mainly the wall echo effects\cite{gibson1978ground}. They have the ability to accurately model the return to isotropy of decaying turbulence and the behavior of turbulence in the rapid distortion limit (\cite{pope2000,mishra4,mishra5,mishra6}). Due to the complex nature of the flow over the AUV form, involving both significant streamline curvature and often flow separation, the use of eddy-viscosity based models may not be optimal for the design and optimization of such structures. Similarly, AUVs operating in deep ocean and river basins often interact with complex turbulence fields because of bed slope and irregular deposition of sediments, which has significant effect on the evolution of turbulent stresses and subsequently on the hydrodynamic parameters such as pressure, skin friction, drag and lift coefficients. At present, there are no studies available in literature in which detailed flow structure along an AUV hull is studied in presence of free stream turbulence with Reynolds stress model based simulations. In this article, the turbulent flow evolution along an AUV hull is studied  both experimentally and numerically, using ADV and high fidelity RSM based simulations for different volumetric Reynolds numbers $(Re_v=\rho U \bigtriangledown^{1/3}/\mu)$, ranging from $0.89\times 10^5$ to $1.31\times 10^5$. The evolution of mean velocity, turbulent kinetic energy and Reynolds stresses are plotted along the AUV hull for different volumetric Reynolds numbers and were used to validate the predictive capability of a Reynolds stress model with the wall reflection term of a pressure strain correlation model. After preliminary validation, The high fidelity Reynolds stress model based simulations were used to predict the hydrodynamic coefficients on the AUV hull with different submergence depths and angle of attacks in both presence and absence of free stream turbulence. This facet of our investigation provides an estimation of the general predictive fidelity of Reynolds Stress Models for the simulation of complex turbulent flows relevant to the design and operation of AUVs.

\section{Experimental setup}
\label{S:2}
\begin{figure}
\centering
\subfloat[Schematic of the recirculating water tank]{\includegraphics[height=6cm]{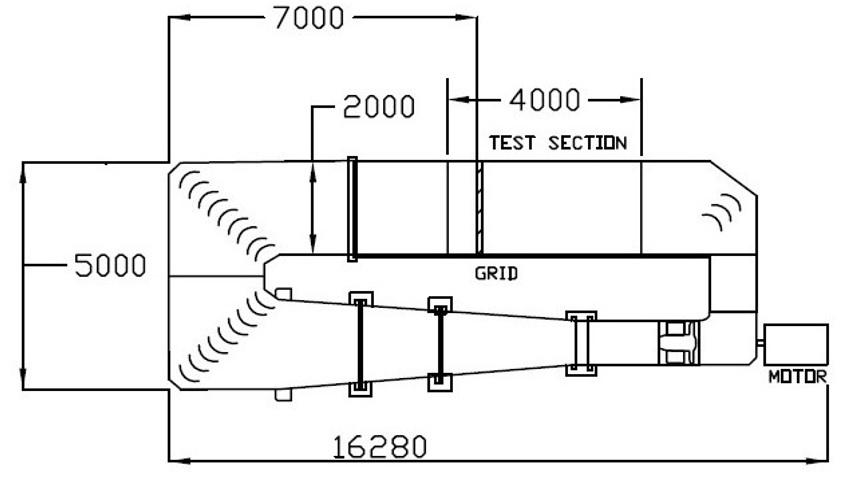}}\\
\subfloat[Experimental setup]{\includegraphics[height=7cm]{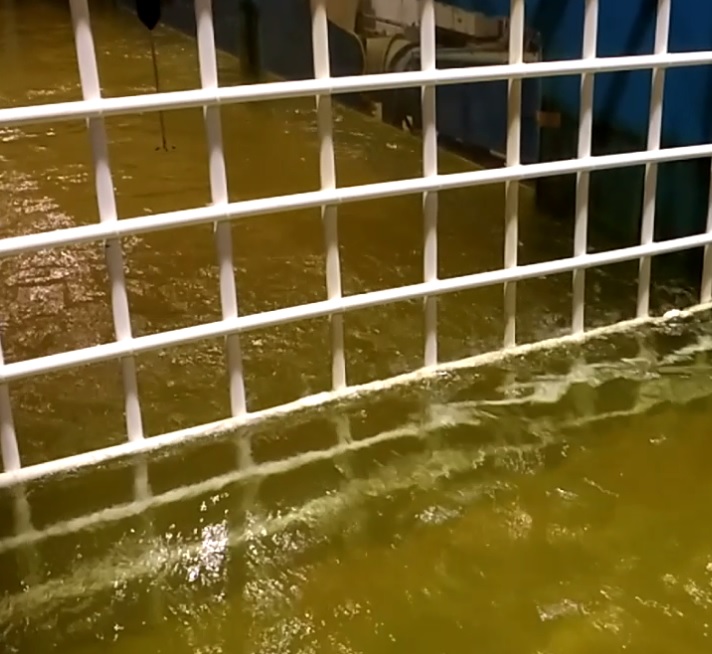}}
\caption{Experimental setup in the recirculating water tank. The schematic diagram is reproduced from \cite{panda2018experimental}. \label{fig:1}}
\end{figure}

The experiments were conducted in a recirculating water tank at the department of Ocean Engineering and Naval Architecture, Indian Institute of Technology, Kharagpur. The water tank side walls are made up of glass. The schematic of the recirculating water tank along with the detailed arrangement of grid, wedge and AUV in the tank is shown in figure \ref{fig:1}a. The water is recirculated by a pump, the rpm of the pump is controlled by an electrical control unit. For a water depth of $0.8$ meter a mean flow velocity up to $1 m/s$ is achievable. The water tank has width 2 meters and depth 1.5 meter. All the experiments were conducted for a water depth of 0.8 meter in the water tank. The experimental setup is shown in figure \ref{fig:1}b. 

The grids made up of cylindrical PVC pipes were placed immediately preceding the test section through a grid holder to produce turbulence as shown in figure \ref{fig:1}b. The mesh length $M$ and diameter of the grid was 0.1 meter and 0.025 meter respectively. More information on the grid used in the experiment is available in \cite{panda2018experimental}.

\begin{figure}
\centering
\subfloat[Only AUV(OA)]{\includegraphics[height=6.2cm]{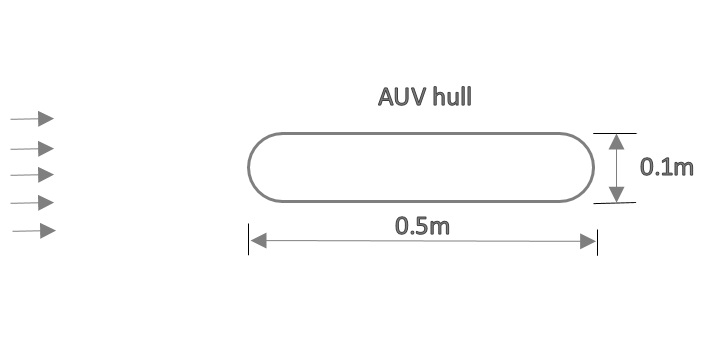}}\\
\subfloat[Turbulence generating Grid and AUV(TA)]{\includegraphics[height=6.5cm]{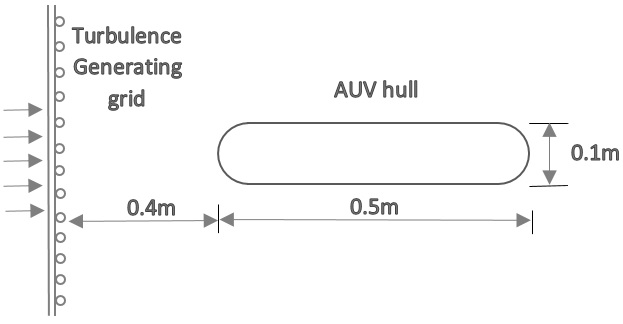}}\\
\caption{Schematic view of the problem geometry, The arrow marks in the figure represents flow direction.\label{fig:111}}
\end{figure}

The AUV hull was fixed at a distance of $0.4$ meter from the grid along the central line of the tank horizontally in the test section of the water tank. A schematic diagram show the setup of the physical flow problem is shown in \ref{fig:111}. The arrows before the AUV hull are representing the flow direction. The AUV hull has a cylindrical body with hemispherical ends as shown in figure \ref{fig:111}. The length and diameter of the AUV hull is 0.5 and 0.1 meter respectively. The diameter of the hemispheres is equal to the diameter of the cylinder. The shape of the AUV hull in this work is based on the geometrical configuration investigated in \cite{mansoorzadeh2014investigation}. Since our main interest is to study the effect of free stream turbulence on flow evolution along the AUV hull, fins were not attached to the hull.

In this setup, the stream wise direction is along the $x$-axis, transverse direction is $y$ axis and $z$ is the vertical direction. The measured instantaneous velocities along the $y$ and $z$ axes are small in comparison to the velocity in $x$ direction. $U, V$ and $W$ are the horizontal, transverse and vertical mean velocities in $x$, $y$ and $z$ directions respectively.

An ADV was used in our experiment to measure instantaneous velocity components at six different locations downstream of the grid across the AUV, which is mostly suitable for flow measurements in  laboratory flumes and hydraulic models with higher sampling rate up to 200 Hz. At each location the ADV was fixed for five minutes, which is sufficient to obtain converged and stable instantaneous velocity data as reported in literature \cite{voulgaris}. A spatial and temporal resolution of $1 {cm^3}$ and $200 Hz$ respectively can be achievable through the ADV used in the experiment.

\begin{figure}
\begin{centering}
\includegraphics[width=0.8\textwidth]{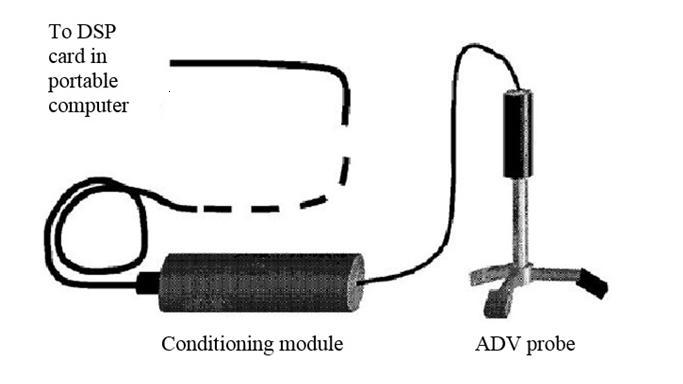}
\caption{Schematic diagram showing ADV probe and signal conditioning module, reproduced from \cite{panda2018experimental}. \label{fig:2}}
\end{centering}
\end{figure}

An ADV can measure instantaneous flow velocities at high sampling rates with very small sampling volume and works on the principle of Doppler shift. The three main components of ADV are a signal conditioning electronic module, sound receivers and sound emitter. The schematic of the acoustic doppler velocimeter is shown in figure \ref{fig:2}. More information on the ADV operation and working principle is available in \cite{panda2018experimental}.

Such experimental data has small errors and uncertainties associated with it, that need to be identified and estimated\cite{benedict1996towards}. The main source of error in the ADV system are the orientation of the probe, local fluid flow properties, velocity range and sampling frequencies. The accuracy of ADV mean flow velocities are with in one percent\cite{voulgaris}. The maximum achievable sampling frequency is 200 Hz, with such high frequency, accuracy of $0.5$ percent of measured value $\pm 1$ mm/s can be achieved for the water velocity. The errors associated with Reynolds stress are with in $1 \%$ of the estimated true value as reported in \cite{voulgaris} for a sampling frequency of 25 Hz.

The experiments were conducted for two different cases in the recirculating water tank. In the first experiment only with AUV flow was investigated and the second one involved the AUV with a turbulence generating grid (hereafter denoted by OA and TA respectively).
\subsection{Data Analysis}
The data collected from the acoustic doppler velocimeter were decomposed into mean and fluctuating velocities in stream-wise, transverse and vertical directions.

The stream-wise mean($U$) and fluctuating($u$) velocities were be calculated from the following formula:
\begin{equation}
U=1/n\sum_{i=1}^{n} U_i
\end{equation}

\begin{equation}
u=\sqrt{1/n\sum_{i=1}^{n} (U_i-U)^2}
\end{equation}
Similar formulas were employed to calculate the mean and fluctuating velocities in other two directions. 

The turbulent kinetic energy is defined as $k=\frac{1}{2}(\overline {u^2} + \overline {v^2} +\overline {w^2})$, which is the mean kinetic energy per unit mass in the fluctuating velocity field. Where v and w are the fluctuating velocities in transverse and vertical directions respectively.
\section{Numerical modeling details}
\label{S:3}
The Reynolds stress transport equation has the form:
\begin{equation}
\begin{split}
\partial_{t} \overline{u_iu_j}+U_k \frac{\partial \overline{u_iu_j}}{\partial x_k}&=P_{ij}-\frac{\partial T_{ijk}}{\partial x_k}-\epsilon_{ij}+\phi_{ij},\\
\mbox{where},\\ 
P_{ij}=-\overline{u_ku_j}\frac{\partial U_i}{\partial x_k}-\overline{u_iu_k}\frac{\partial U_j}{\partial x_k}, & T_{kij}=\overline{u_iu_ju_k}-\nu \frac{\partial \overline{u_iu_j}}{\partial{x_k}}+\delta_{jk}\overline{ u_i \frac{p}{\rho}}+\delta_{ik}\overline{ u_j \frac{p}{\rho}}\\
,\epsilon_{ij}=-2\nu\overline{\frac{\partial u_i}{\partial x_k}\frac{\partial u_j}{\partial x_k}},& \phi_{ij}= \overline{\frac{p}{\rho}(\frac{\partial u_i}{\partial x_j}+\frac{\partial u_j}{\partial x_i})}\\
\end{split}
\end{equation}
$P_{ij}$ denotes the production of turbulence, $T_{ijk}$ is the diffusive transport, $\epsilon_{ij}$ is the dissipation rate tensor and $\phi_{ij}$ is the pressure strain correlation. The pressure fluctuations are governed by a Poisson equation:
\begin{equation}
\frac{1}{\rho}{\nabla}^2(p)=-2\frac{\partial{U}_j}{\partial{x}_i}\frac{\partial{u}_i}{\partial{x}_j}-\frac{\partial^2 u_iu_j}{\partial x_i \partial x_j}
\end{equation}

The fluctuating pressure term is split into a slow and rapid pressure term $p=p^S+p^R$. Slow and rapid pressure fluctuations satisfy the following equations
\begin{equation}
\frac{1}{\rho}{\nabla}^2(p^S)=-\frac{\partial^2}{\partial x_i \partial x_j}{(u_iu_j-\overline {u_iu_j})}
\end{equation}
\begin{equation}
\frac{1}{\rho}{\nabla}^2(p^R)=-2\frac{\partial{U}_j}{\partial{x}_i}\frac{\partial{u}_i}{\partial{x}_j}
\end{equation}
It can be seen that the slow pressure term accounts for the non-linear interactions in the fluctuating velocity field and the rapid pressure term accounts for the linear interactions. A general solution for $\phi_{ij}$ can be obtained by applying Green's theorem to equation (7):
\begin{equation}
\phi_{ij}=\frac{1}{4\pi}\int_{-\infty}^{\infty} \overline{\frac{\partial{u}^*_k}{\partial{x_l^*}}\frac{\partial{u}^*_l}{\partial{x_k^*}}(\frac{\partial{u}_i}{\partial{x_j}}+\frac{\partial{u}_j}{\partial{x_i}})}
+2G_{kl}\overline{\frac{\partial{u}^*_l}{\partial{x_k^*}}(\frac{\partial{u}_i}{\partial{x_j}}+\frac{\partial{u}_j}{\partial{x_i}})}\frac{dVol^*}{|{x_n-x_n^*}|}
\end{equation}
The volume element of the corresponding integration is $dVol^*$. Instead of an analytical approach, the pressure strain correlation is modeled using rational mechanics approach. The rapid term can be modeled by assuming the length scale of mean velocity gradient is much larger than the turbulent length scale and is written in terms of a fourth rank tensor \cite{pope2000}
\begin{equation}
\phi_{ij}^R=4k\frac{\partial{U}_l}{\partial{x_k}}(M_{kjil}+M_{ikjl})
\end{equation}
where, 
\begin{equation}
M_{ijpq}=\frac{-1}{8\pi k}\int \frac{1}{r} \frac {\partial^2 R_{ij}(r)}{\partial r_p \partial r_p}dr
\end{equation}
where, $R_{ij}(r)=\langle u_i(x)u_j(x+r) \rangle$

For homogeneous turbulence the complete pressure strain correlation can be written as
\begin{equation}
\phi_{ij}=\epsilon A_{ij}(b)+kM_{ijkl}(b)\frac{\partial\overline {v}_k}{\partial{x_l}}
\end{equation}
The most general form of slow pressure strain correlation is given by
\begin{equation}
\phi^{S}_{ij}=\beta_1 b_{ij} + \beta_2 (b_{ik}b_{kj}- \frac{1}{3}II_b \delta_{ij})
\end{equation}
Established slow pressure strain correlation models including the models of \cite{rotta1951} and \cite{ssmodel} use this general expression. Considering the rapid pressure strain correlation, the linear form of the model expression is
\begin{equation}
\frac{\phi^{R}_{ij}}{k}=C_2 S_{ij} +C_3 (b_{ik}S_{jk}+b_{jk}S_{ik}-\frac{2}{3}b_{mn}S_{mn}\delta_{ij})+  
C_4 (b_{ik}W_{jk} + b_{jk}W_{ik})
\end{equation}
Here $b_{ij}=\frac{\overline{u_iu_j}}{2k}-\frac{\delta_{ij}}{3}$ is the Reynolds stress anisotropy tensor, $S_{ij}$ is the mean rate of strain and $W_{ij}$ is the mean rate of rotation. Rapid pressure strain correlation models like the models of \cite{mishra6} use this general expression.

The wall reflection term, redistributes the normal stress near the wall and it damps the component of Reynolds stress which is perpendicular to the wall and enhances the Reynolds stress parallel to the wall as reported in \cite{pope2000}. The wall reflection has both slow and rapid contributions and can be written as:
\begin{equation}
\begin{split}
\phi_{ij,Sw}=0.5\frac{\epsilon}{k}\bigg(\overline{u_{k}u_{m}}n_{k}n_{m}\delta{ij}-1.5\overline{u_{i}u_{k}}n_{j}n_{k}-1.5\overline{u_{j}u_{k}}n_{i}n_{k}\bigg)\frac{c_{l}k^{3/2}}{\epsilon d}\\
\phi_{ij,Rw}=0.3\phi_{km,R}\frac{\epsilon}{k}\bigg(n_{k}n_{m}\delta{ij}-1.5\phi_{ik,R}n_{j}n_{k}-1.5\phi_{jk,R}n_{i}n_{k}\bigg)\frac{c_{l}k^{3/2}}{\epsilon d}
\end{split}
\end{equation}
where, $n_k$ and $x_k$ are the components of the unit normal to the wall, $d$ is the normal distance to the wall\cite{gibson1978ground,ansys}.

The computational fluid dynamics(CFD) simulations were performed using the ANSYS Fluent solver \cite{ansys}. The incompressible Navier-Stokes equations were solved utilizing a control volume based method. The velocity and pressure fields were coupled using the SIMPLE scheme (Semi-Implicit Method for Pressure Linked Equations). The model was generated using GAMBIT meshing utility and create spatial meshes predominantly consisting of tetrahedral elements. For mesh independence 3 different meshes of increasing resolution were generated. These had $0.7 \times 10^6$, $1.4 \times 10^6$ and $2.1 \times 10^6$ cells respectively. In all these Inflation layers up till 5 cells from all walls were utilized to ensure that that $y^+$ were adequate for wall functions to be utilized. From the numerical simulations it was observed that the second and third mesh predicts almost identical turbulent flow field along the AUV hull with negligible difference. For all the simulations we have used the mesh with 2.1 million cells. The inlet and outlet of the water tank were modeled as a velocity inlet with a uniform inflow and a pressure outlet respectively.

\section{Results and discussions}
\label{S:4}
\subsection{Experimental results}
The measurement of instantaneous three dimensional velocities were taken along the length of the AUV at a distance of 0.05 m from the side walls and at six equidistant points starting from the beginning towards the end of the AUV hull. The mean velocity measurement is non-dimensionalised by U (the time averaged free stream mean velocity), the stresses and turbulent kinetic energy are non-dimensionalised by $U^2$.

\begin{figure}
\centering
 \subfloat[$Re_v=0.89\times 10^5$]{\includegraphics[width=0.55\textwidth]{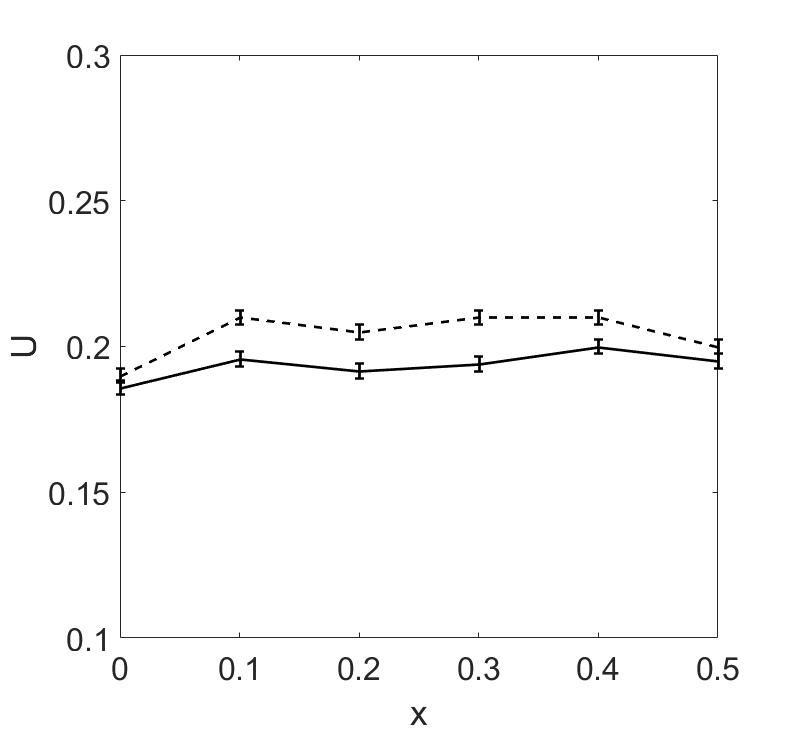}}
\subfloat[$Re_v=1.11\times 10^5$]{\includegraphics[width=0.55\textwidth]{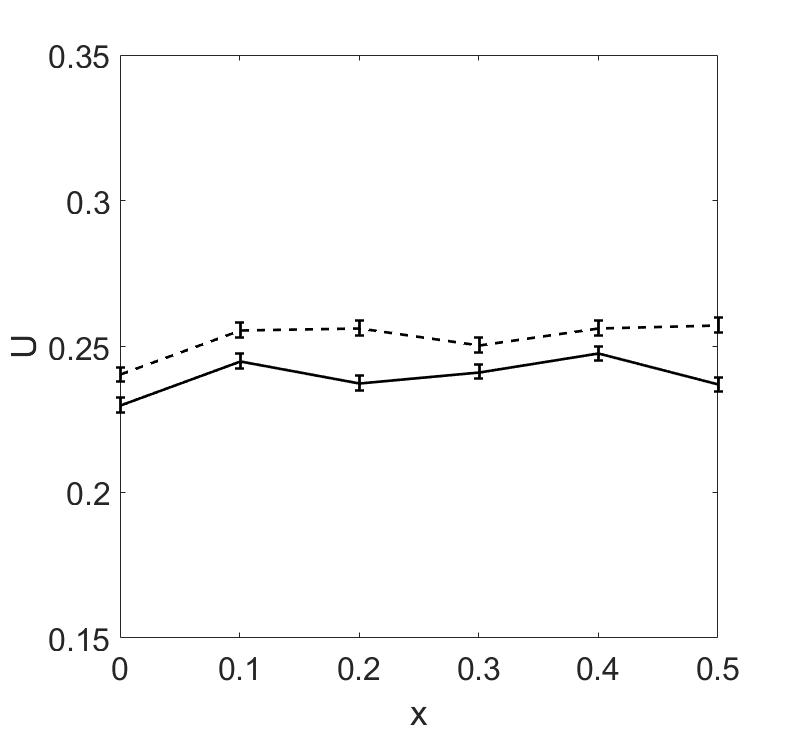}}\\
\subfloat[$Re_v=1.31\times 10^5$]{\includegraphics[width=0.55\textwidth]{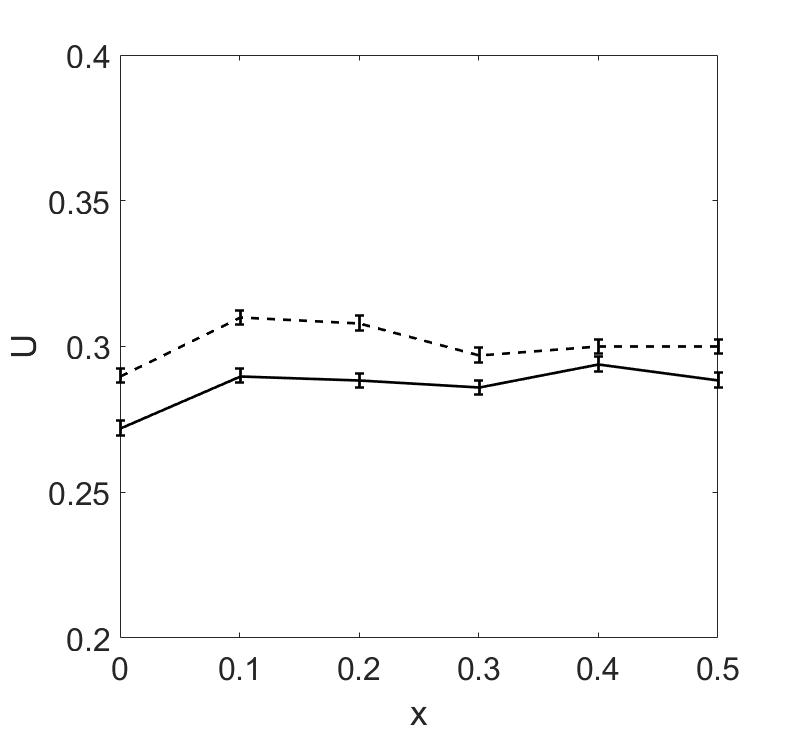}}
 \caption{Experimental results of mean velocity along the AUV hull with error bars for three different volumetric Reynolds numbers, dotted lines for only AUV(OA) and solid lines for turbulence and AUV(TA) case. \label{fig:3}}
\end{figure}

In figure \ref{fig:3}, the profile of mean velocity is shown for both OA and TA cases for three different grid Reynolds numbers. The error bars were also included in the figures to accommodate the uncertainties in the measurements.
\begin{figure}
\centering
 \subfloat[$Re_v=0.89\times 10^5$]{\includegraphics[width=0.55\textwidth]{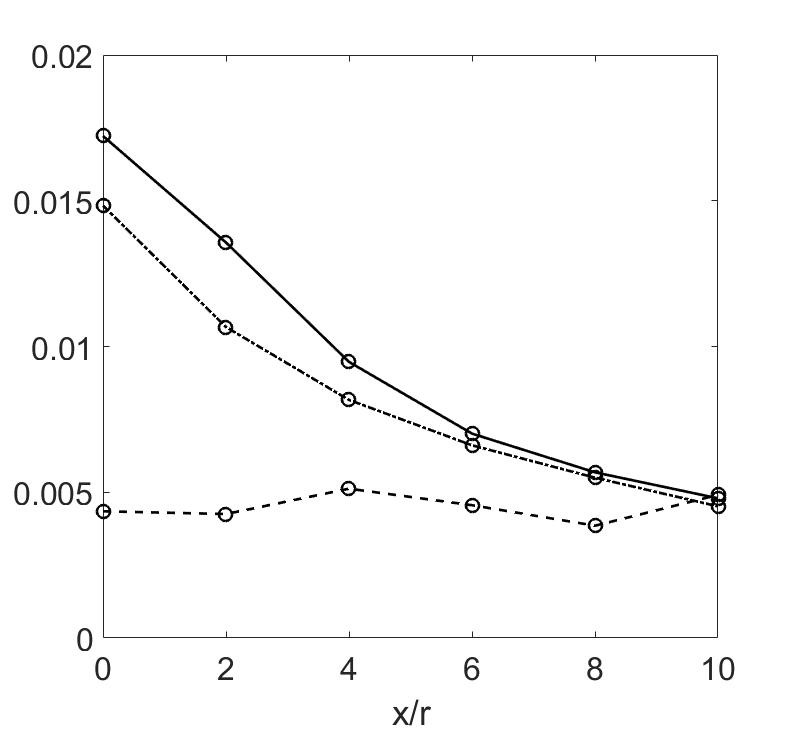}}
 \subfloat[$Re_v=1.11\times 10^5$]{\includegraphics[width=0.55\textwidth]{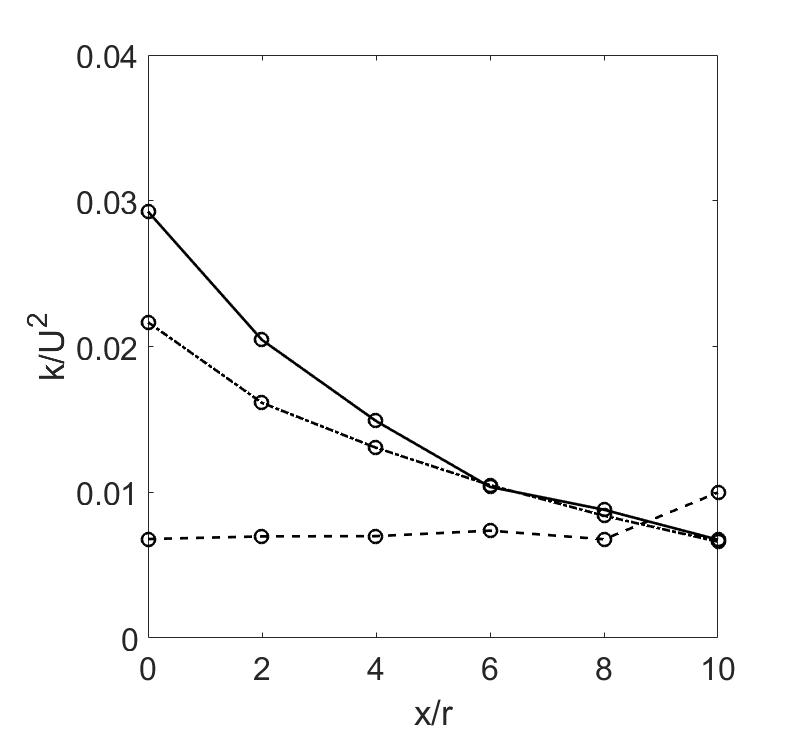}}\\
 \subfloat[$Re_v=1.31\times 10^5$]{\includegraphics[width=0.55\textwidth]{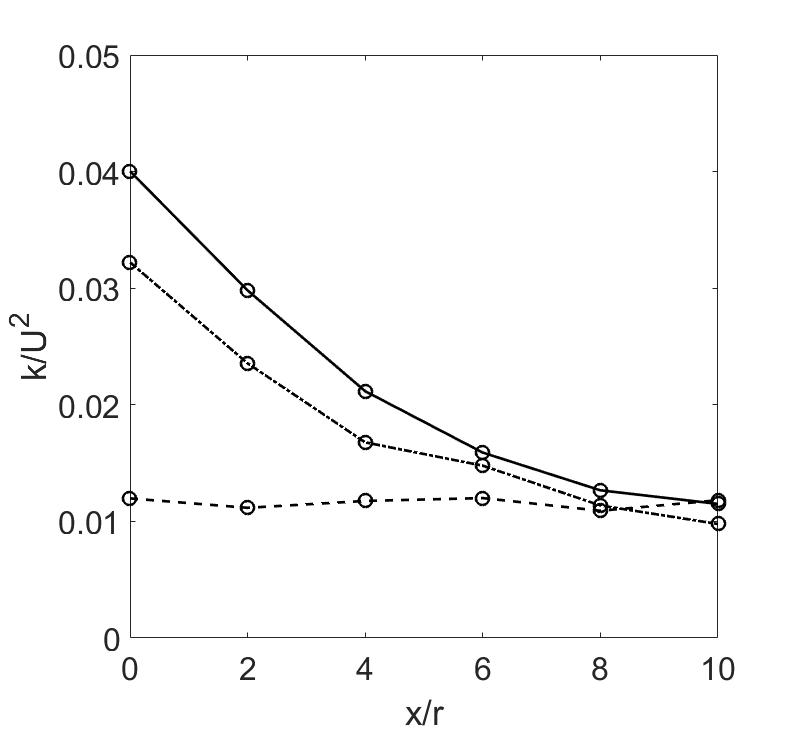}}
 \caption{Experimental results of evolution of turbulence kinetic energy along the AUV hull for three different volumetric Reynolds numbers. Dashed line for only AUV case(OA), Dashed-dot line for only grid (decaying turbulence) case from \cite{panda2018experimental}, solid line for grid turbulence and AUV(TA) case. \label{fig:4}}
\end{figure}

Figure \ref{fig:4} shows the variation of turbulence kinetic energy along the AUV hull for three volumetric Reynolds numbers. Experimental results of decaying free stream turbulence (dashed-dot line) were included from Panda et al.\cite{panda2018experimental} to compare the experimental results of OA and TA cases. It is observed that in presence of the turbulence generating grid the TKE level along the AUV hull increases.  

\subsection{Validation of the numerical model}
\begin{figure}
\centering
 \subfloat[ ]{\includegraphics[width=0.5\textwidth]{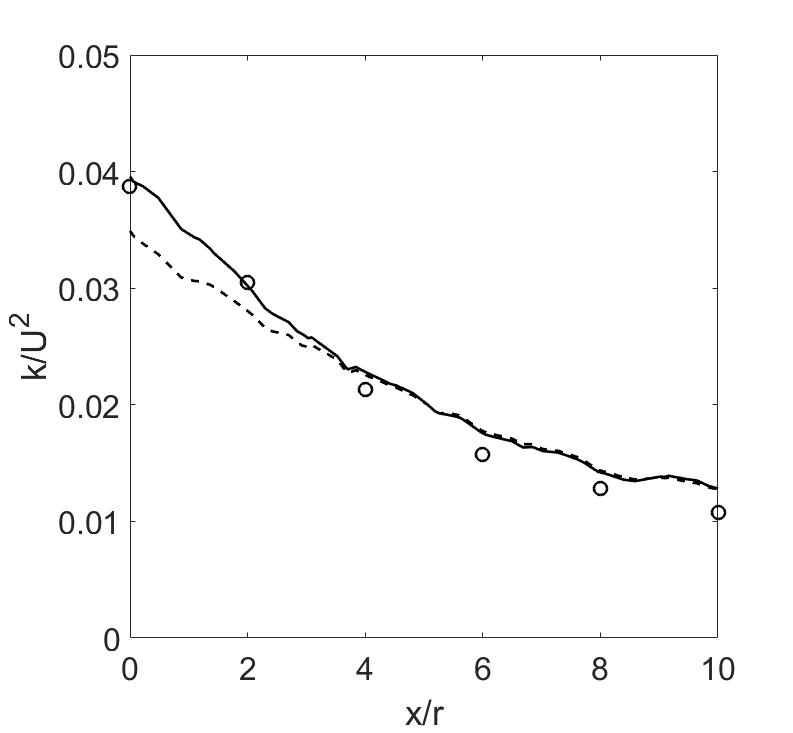}}
 \subfloat[ ]{\includegraphics[width=0.5\textwidth]{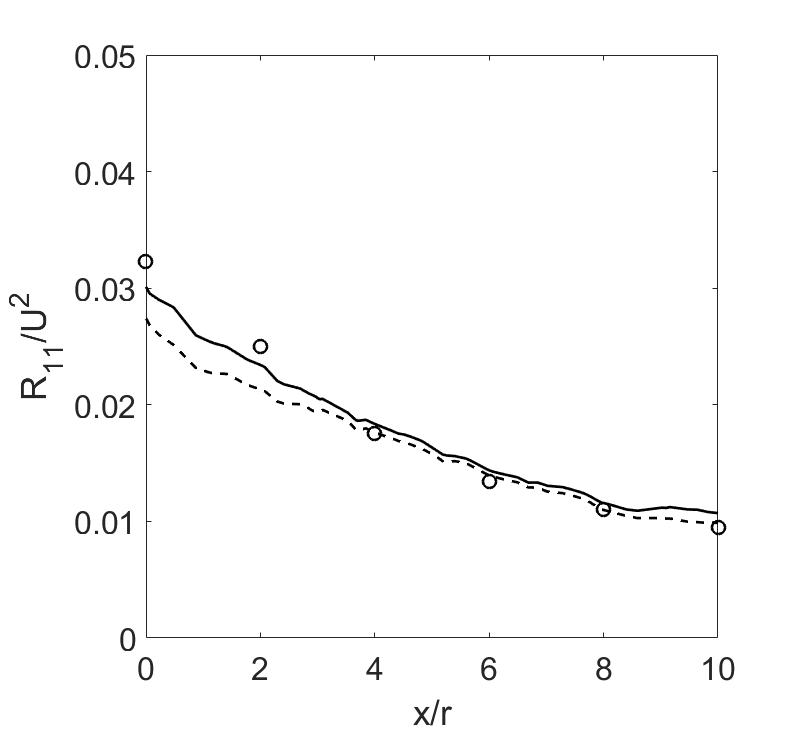}}\\
 \subfloat[ ]{\includegraphics[width=0.5\textwidth]{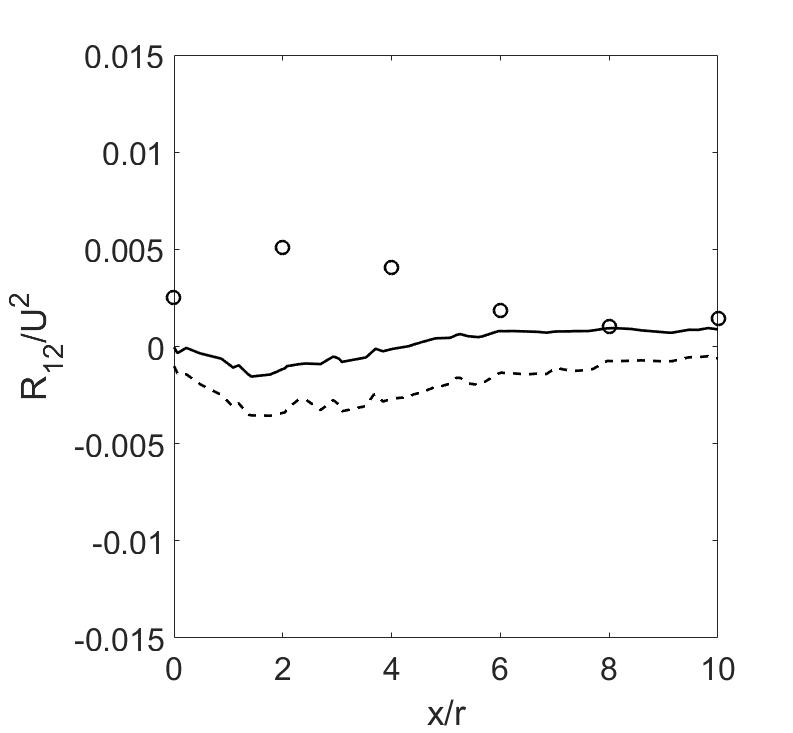}}
 \subfloat[ ]{\includegraphics[width=0.5\textwidth]{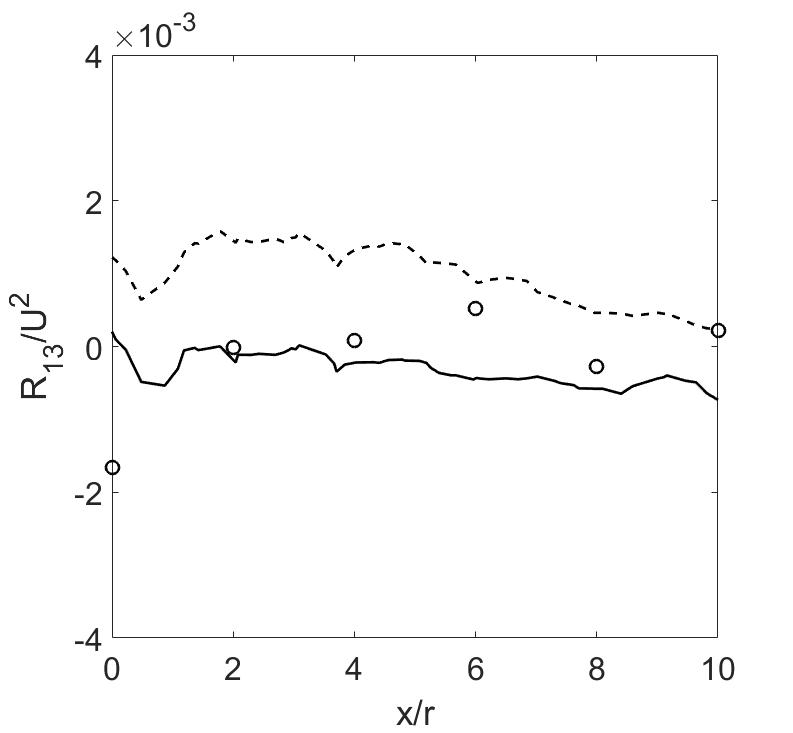}}
 \caption{Comparison of the Reynolds stress model predictions with the turbulence statistics along the AUV hull in presence of free stream turbulence, a) TKE evolution, b) evolution of normal component of Reynolds stress, c) and d) evolution of shear components of Reynolds stress. Dashed lines Case 1(without wall reflection term) and Solid lines for case 2 (with wall reflection term  of the pressure strain correlation model). \label{fig:5}}
\end{figure}

For the validation of the numerical model, the experimental results for the case of flow past AUV in presence of grid is considered. The numerical simulations were conducted with the linear pressure strain model. The detailed formulation of the linear pressure strain model is available in \cite{ansys}. We have considered two different cases for our numerical simulations. For case 1, The simulations were conducted only with the slow and rapid component of the pressure strain model and for Case 2, In addition to slow and the rapid term the wall reflection term was also considered.  Figure \ref{fig:5} shows the comparison of the pressure strain correlation model predictions of the turbulence kinetic energy and the components of the Reynolds stresses (both normal and shear). From all the figures it is clear that, Case 2 predicts better results in comparison to Case 1, this is because of the addition of wall reflection term to the pressure strain correlation modeling basis\cite{gibson1978ground}. From figure\ref{fig:5}b it is clear that, the wall reflection term accurately captures the modified pressure field in the proximity of the rigid AUV wall and impede the transfer of energy from the stream wise direction to that normal to the wall. So for all the simulations in this paper the pressure strain model with the wall reflection term (Case 2) is considered.

\subsection{Numerical results}

The drag coefficient is given by:
 \begin{equation}
 C_d=\frac{F_d}{0.5\rho U^2A}
 \end{equation}
 Where, $F_d$ is the drag force acting on the cylinder, $A$ is the area of the external surface of the model.
 
 The periodic fluctuations of the flow exerts certain force on the AUV hull and that is characterized by the lift coefficient:
 \begin{equation}
 C_L=\frac{F_L}{0.5\rho U^2A}
 \end{equation}
 
 The pressure coefficient is defined as:
 \begin{equation}
 C_p=\frac{p-p_\infty}{0.5\rho U^2}
 \end{equation}
 Where, p is the pressure at the point for which pressure coefficient is calculated.
 
\subsubsection{Effect of FST on hydrodynamic parameters at different Reynold numbers}
\begin{figure}
\centering
 \subfloat[ ]{\includegraphics[width=0.5\textwidth]{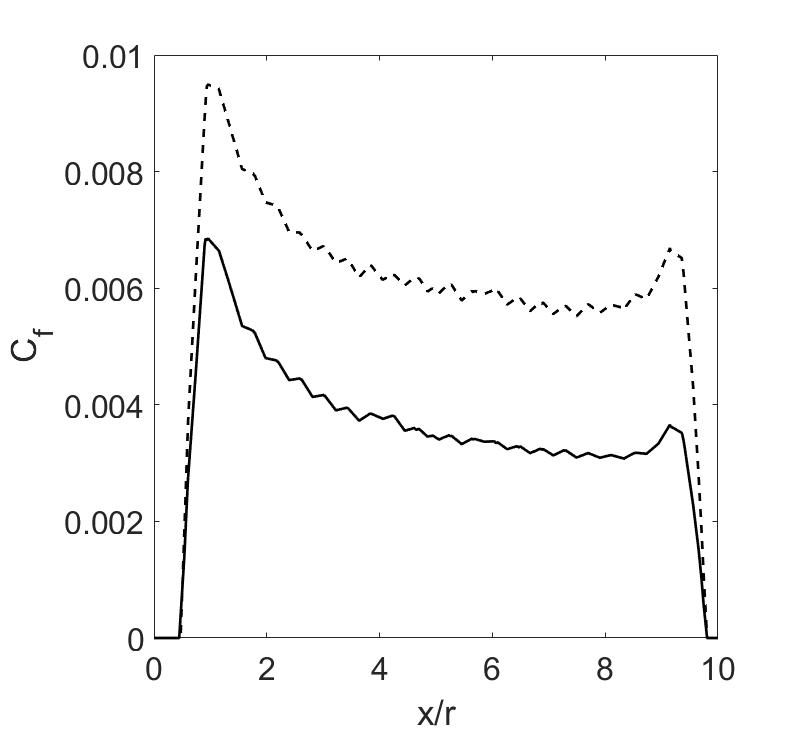}}
 \subfloat[ ]{\includegraphics[width=0.5\textwidth]{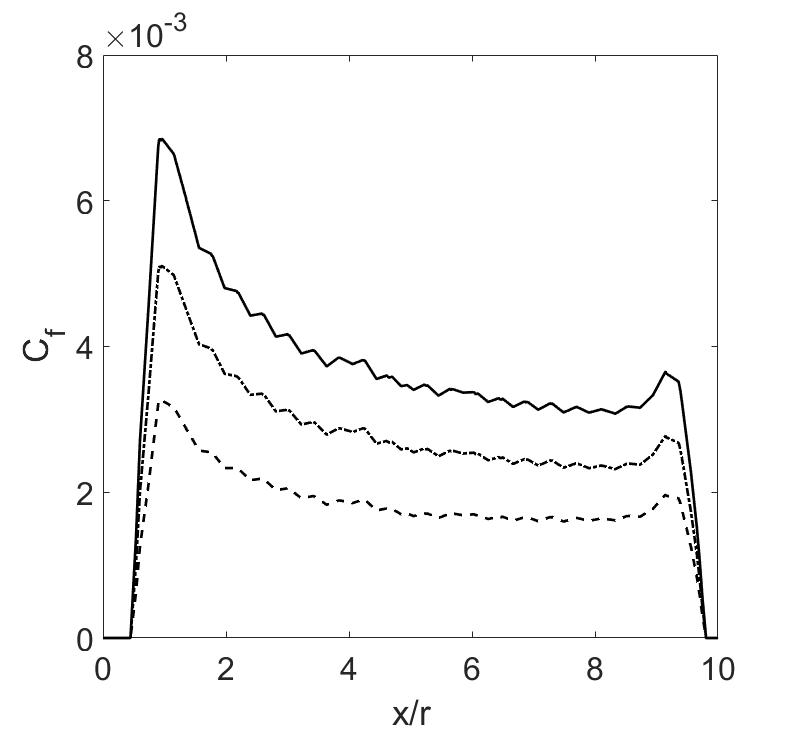}}\\
 \caption{Turbulence skin friction evolution on the wall of the AUV along its length away from the grid (a) comparison of OA (dashed line) and TA (solid line) cases for $Re_v=1.31\times 10^5$ b) comparison of $C_f$ evolution for TA case for different levels of free stream turbulence, solid line $Re_v=1.31\times 10^5$, dashed dot line $Re_v=1.11\times 10^5$ and dashed line $Re_v=0.89\times 10^5$.\label{fig:6}}
\end{figure}

\begin{figure}
\centering
 \subfloat[ ]{\includegraphics[width=0.5\textwidth]{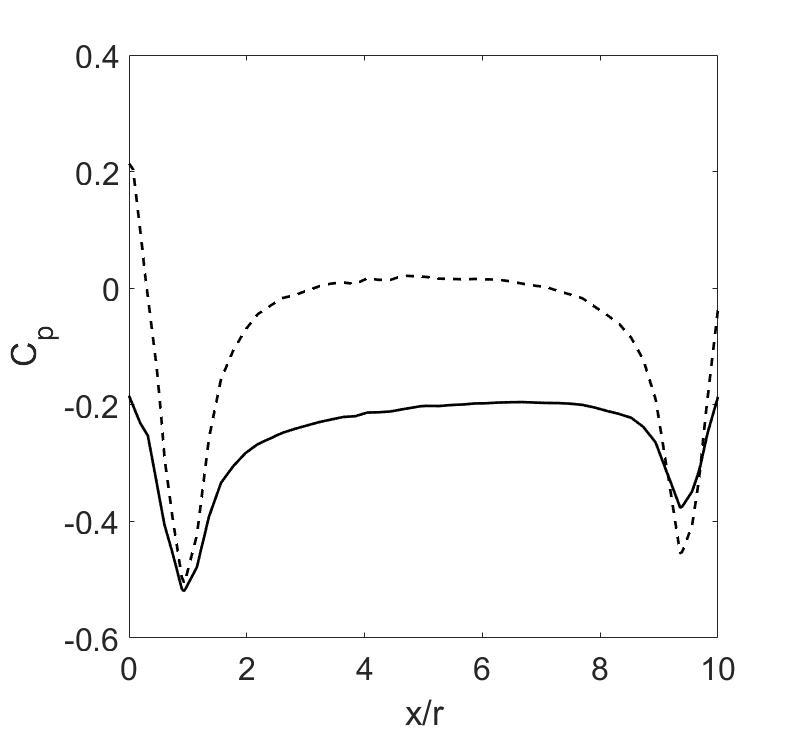}}
 \subfloat[ ]{\includegraphics[width=0.5\textwidth]{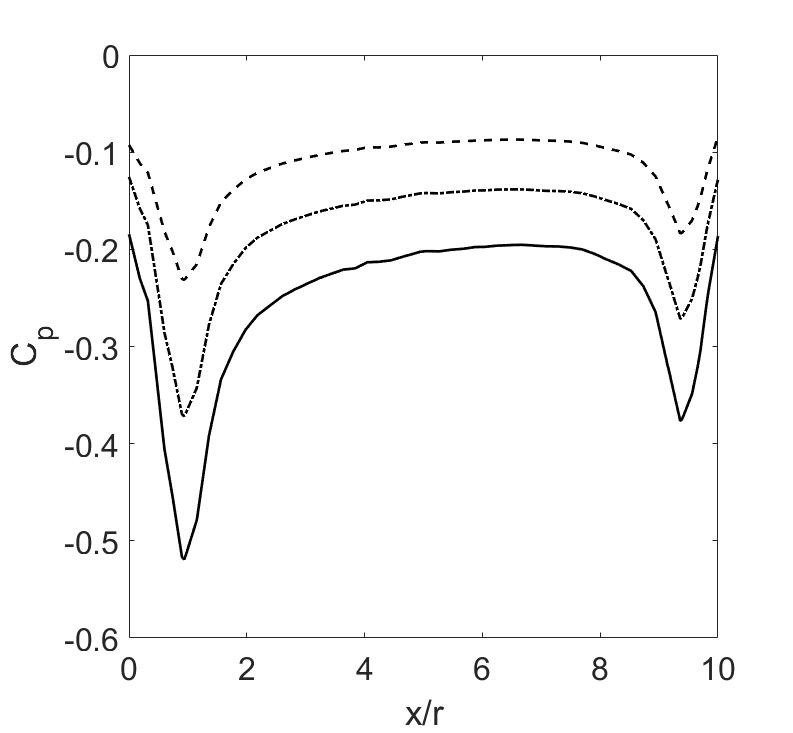}}\\
 \caption{Pressure coefficient evolution on the wall of the AUV along its length away from the grid (a) comparison of OA (dashed line) and TA (solid line) cases for $Re_v=1.31\times 10^5$ b) comparison of $C_p$ evolution for TA case for different levels of free stream turbulence, solid line $Re_v=1.31\times 10^5$, dashed dot line $Re_v=1.11\times 10^5$ and dashed line $Re_v=0.89\times 10^5$.\label{fig:7}}
\end{figure}

\begin{figure}
\centering
\subfloat[ ]{\includegraphics[width=0.5\textwidth]{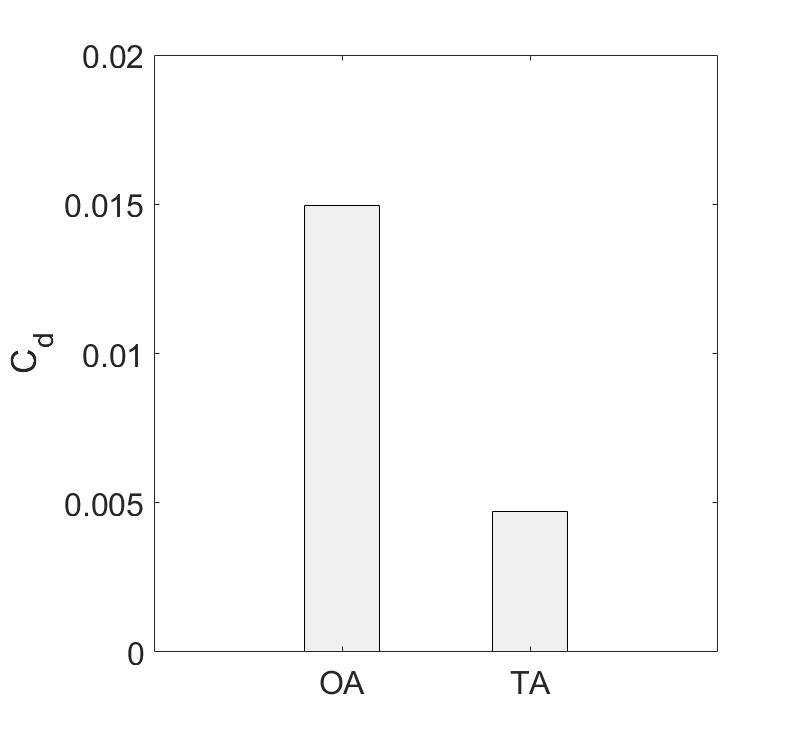}}
 \subfloat[ ]{\includegraphics[width=0.5\textwidth]{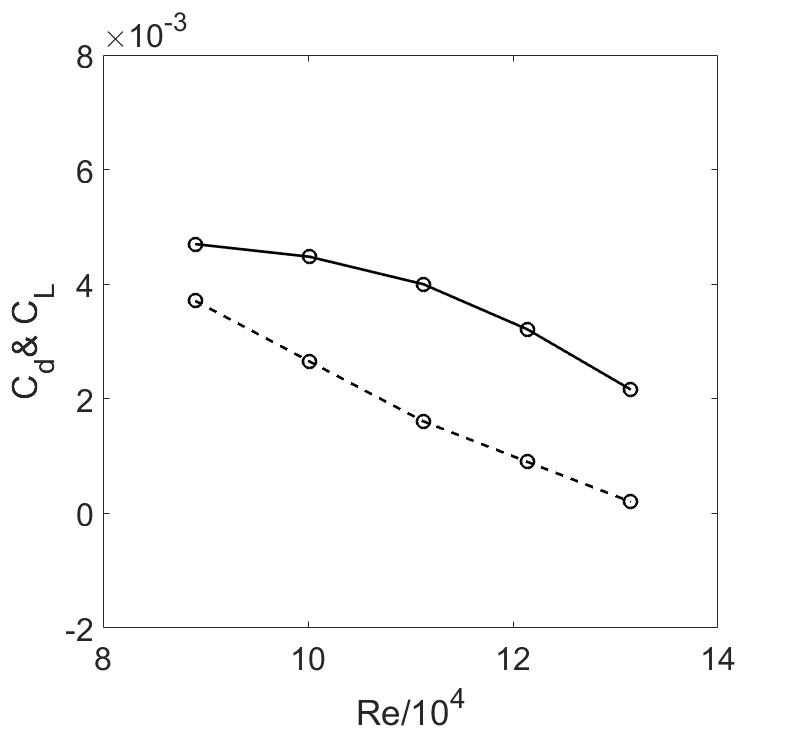}}\\
 \caption{Effect of free stream turbulence on drag and lift coefficients of the AUV, a) bar graph representation of drag coefficient for OA and TA case for $Re_v=1.31\times 10^5$ b) variation of drag and lift coefficients for the TA case for different volumetric Reynolds numbers \label{fig:8}}
\end{figure}

figure \ref{fig:6} depicts the variation of hydrodynamic parameters at different volumetric Reynolds numbers($Re_v$) for both OA and TA cases. From figure \ref{fig:6}a it is clear that in presence of free stream turbulence the skin friction coefficient decreases on the AUV hull along its length. The variation of skin friction coefficient for three different $Re_v$ is presented in figure \ref{fig:6}b for the case of AUV with grid turbulence. A similar trend was observed for the evolution of pressure coefficient along the AUV hull as shown in figure \ref{fig:7}. With increase in turbulence kinetic energy a decrease in pressure coefficient is observed. In figure \ref{fig:8} a comparison of drag coefficients of the AUV for OA and TA cases is shown. From figure \ref{fig:8}a it is observed that free stream turbulence decreases drag coefficient by $68.5\%$ for $Re_v=1.31\times 10^5$, which is in accordance with the findings of \cite{son2010effect} in which they have studied the effects of free stream turbulence on flow over a sphere and reported a drag reduction of $70\%$. figure \ref{fig:8}b represents the variation of drag and lift coefficients for different volumetric Reynolds numbers for TA case. It is noticed that with increase in $Re_v$ or conversely with increase in turbulence kinetic energy both $C_d$ and $C_L$ of the AUV decreases. As reported in literature free stream turbulence suppresses the strength of vortex shedding and reduces the length of recirculation zone behind a bluff body  \cite{mujumdar1970eddy,bakic2003experimental}, and results in drag reduction with increase in turbulence kinetic energy\cite{moradian2009effects}.

\begin{figure}
\centering
 \subfloat[ ]{\includegraphics[width=0.5\textwidth]{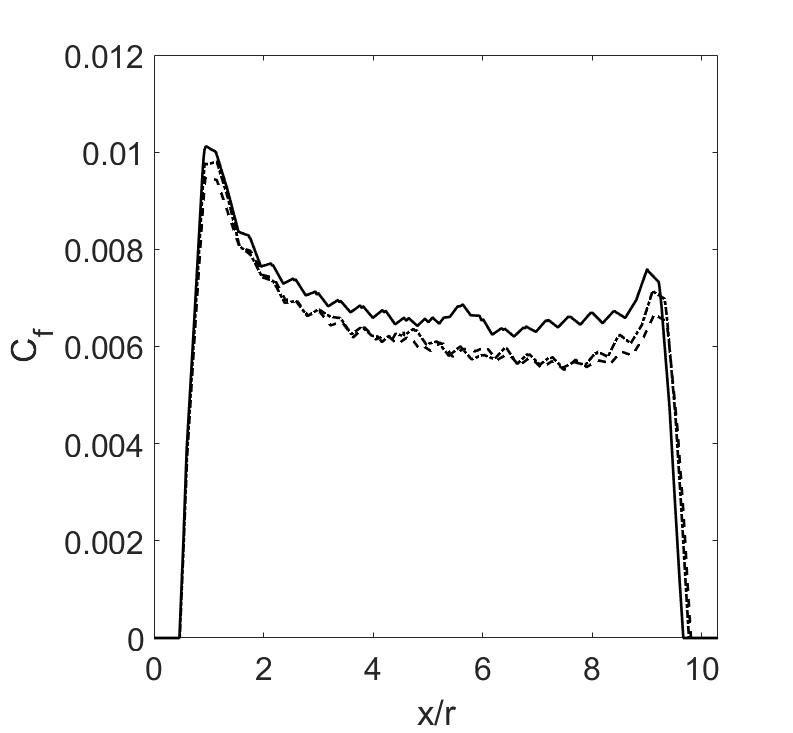}}
 \subfloat[ ]{\includegraphics[width=0.5\textwidth]{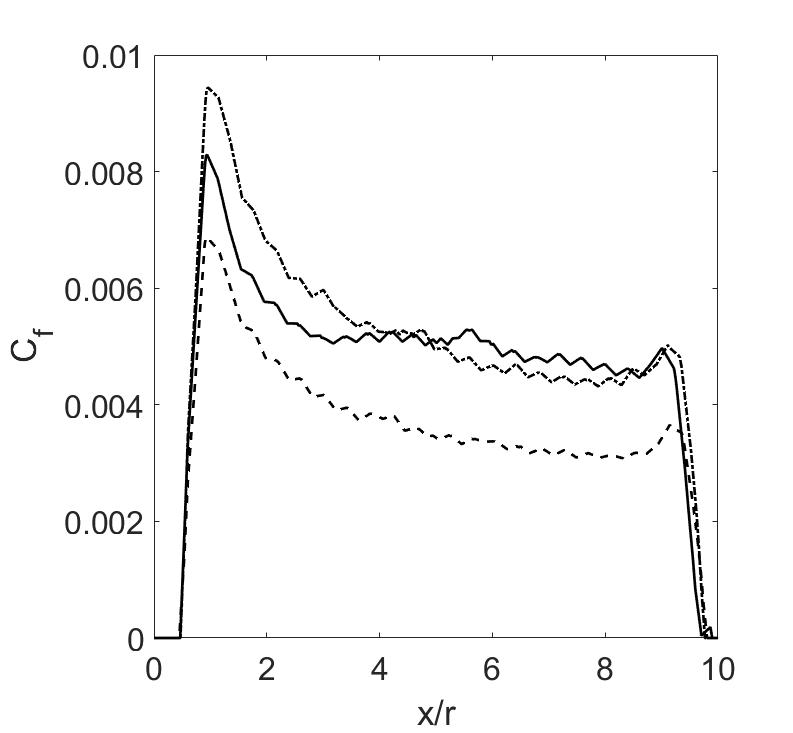}}\\
 \caption{Skin friction coefficient evolution on the wall of the AUV along its length away from the grid. Comparison of numerical results for different value of angle of attack. dashed line 0 degree, dashed dot line 5 degree and solid line 10 degree. a) Only AUV(OA) case b) Turbulence and AUV (TA) case.\label{fig:9}}
\end{figure}

\begin{figure}
\centering
 \subfloat[ ]{\includegraphics[width=0.5\textwidth]{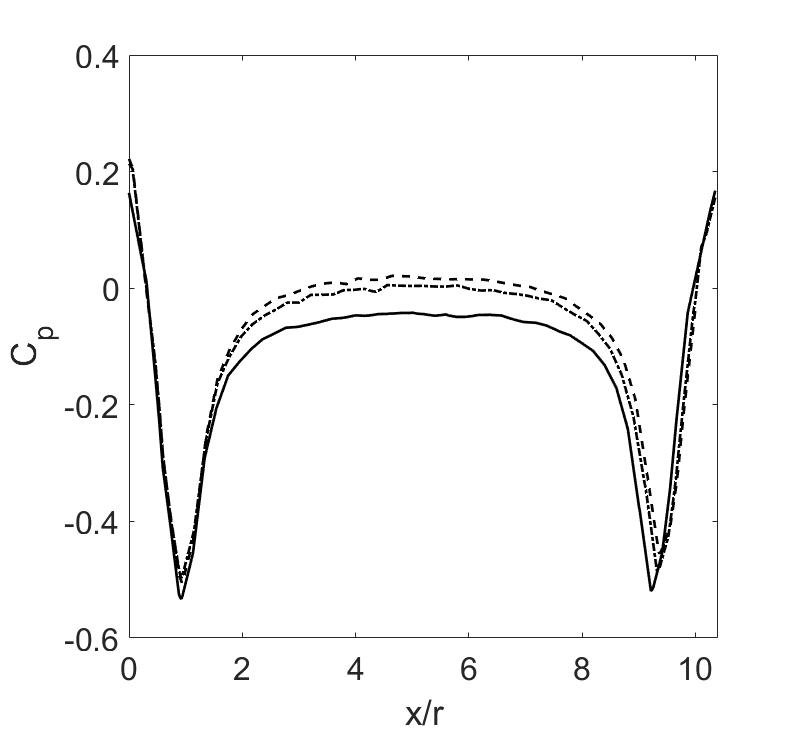}}
 \subfloat[ ]{\includegraphics[width=0.5\textwidth]{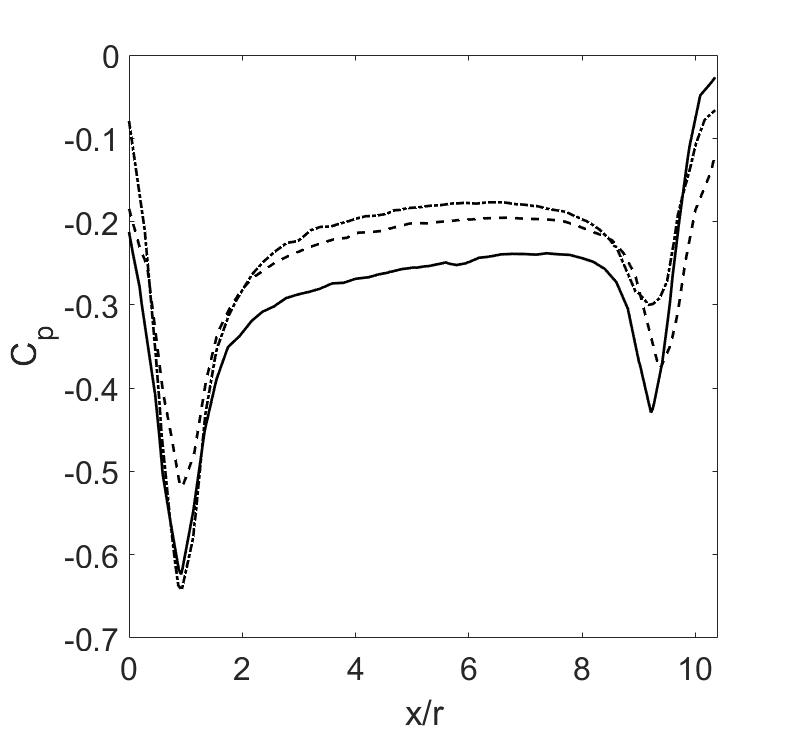}}\\
 \caption{Pressure coefficient evolution on the wall of the AUV along its length away from the grid. Comparison of numerical results for different value of angle of attack. dashed line 0 degree, dashed dot line 5 degree and solid line 10 degree. a) Only AUV(OA) case b) Turbulence and AUV (TA) case.\label{fig:10}}
\end{figure}

\begin{figure}
\centering
 \subfloat[ ]{\includegraphics[width=0.5\textwidth]{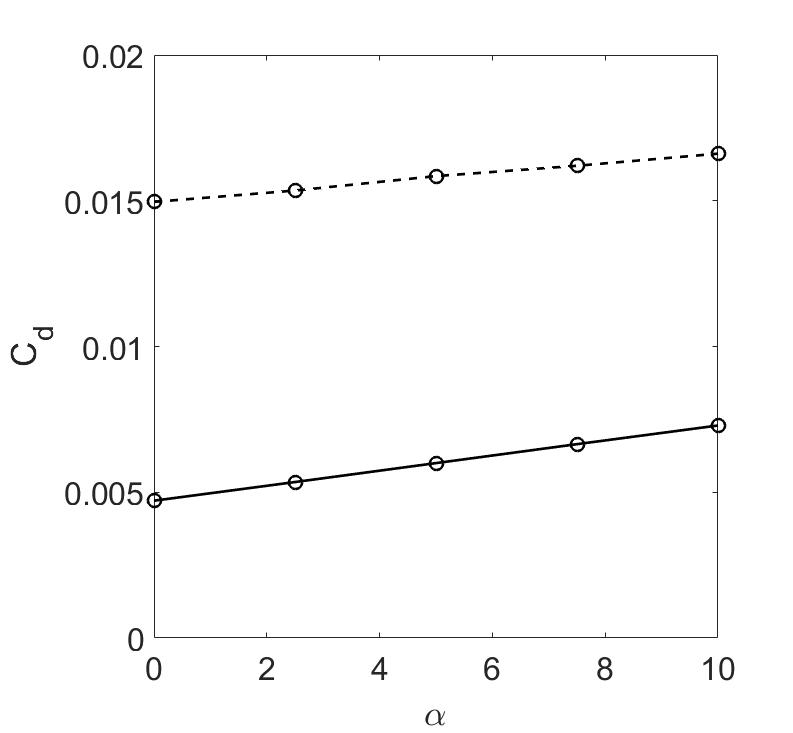}}
 \subfloat[ ]{\includegraphics[width=0.5\textwidth]{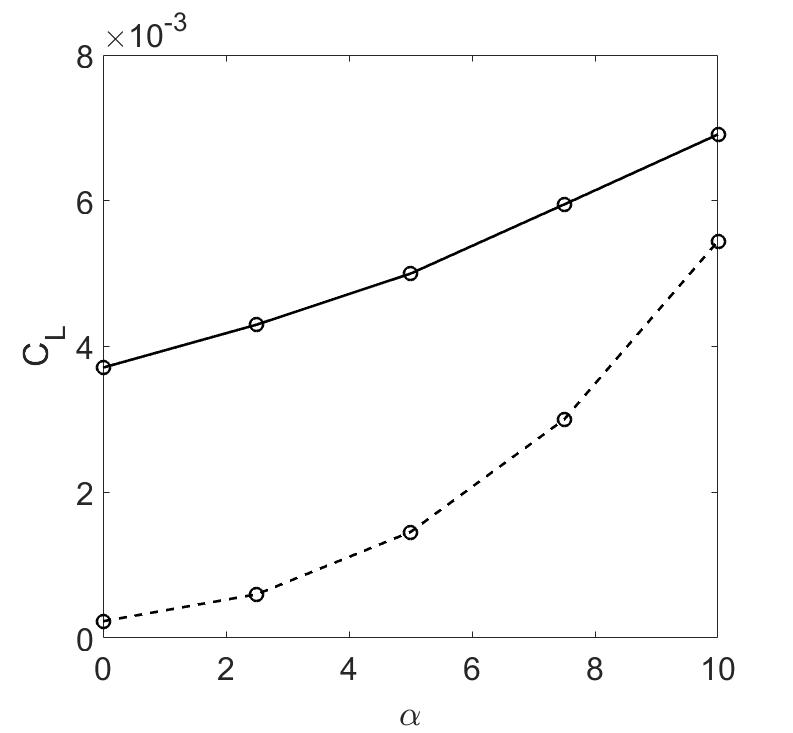}}\\
 \caption{Effect of angle of attack on drag and lift coefficients of the AUV, a) Drag coefficient b) Lift Coefficient. Dashed line OA case and solid line TA case.\label{fig:11}}
\end{figure}

\subsubsection{Effect of FST on hydrodynamic parameters at different angle of attacks}

figure \ref{fig:9}, \ref{fig:10} and \ref{fig:11}  show the variation of $C_f$, $C_p$, $C_d$ and $C_L$ for different values of angle of attacks both for the cases of OA and TA for $Re_v=1.31\times 10^5$. In figure \ref{fig:9} the variation of skin coefficient along the length of the AUV hull for three different values of angle of attacks is presented. Figure \ref{fig:9}a presents the $c_f$ evolution for OA case. It is observed that with increase in angle of attack the skin friction coefficient increases. A larger magnitude of $C_f$ is observed towards the end of the AUV hull. The $C_f$ values further increases with increase in volumetric Reynolds numbers in presence of the turbulence generating grid as shown in figure \ref{fig:9}b. figure \ref{fig:10} represents the evolution of pressure coefficient along the AUV hull for different AOA. A reverse trend is observed for the evolution of $C_p$ in contrast to the evolution of $C_f$ with AOA variation. With increase in AOA the drag and lift coefficient increases along the AUV hull as shown in figure \ref{fig:11} for the OA case, a similar trend was also observed for the evolution of $C_d$ and $C_l$ in the experiments of \cite{jagadeesh2009experimental}. As observed from figure \ref{fig:9}a the free stream turbulence turbulence reduces the $C_d$ at all five angle of attacks. In figure \ref{fig:9}b the variation of $C_L$ is presented for five AOA. It is noticed that with increase in AOA lift coefficient increases.  

\subsubsection{Effect of FST on hydrodynamic parameters at different submergence depths}
\begin{figure}
\centering
 \subfloat[ ]{\includegraphics[width=0.5\textwidth]{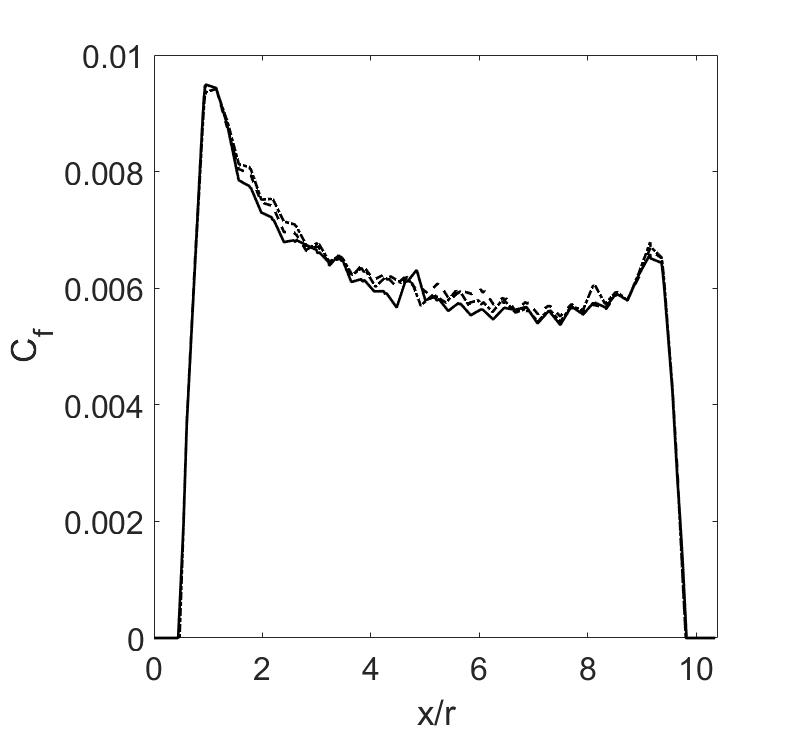}}
 \subfloat[ ]{\includegraphics[width=0.5\textwidth]{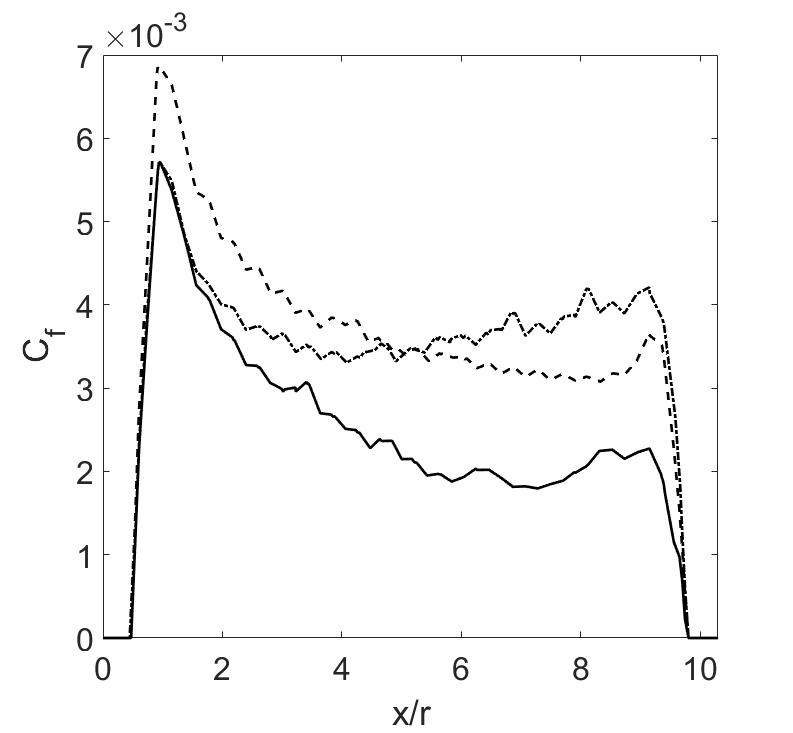}}\\
 \caption{Skin friction coefficient evolution on the wall of the AUV along its length away from the grid. Comparison of numerical results for different value of submergence depth. dashed line $h/r=0$, dashed dot line $h/r=2$ and solid line $h/r=4$. a) Only AUV(OA) case b) Turbulence and AUV (TA) case.\label{fig:12}}
\end{figure}

\begin{figure}
\centering
 \subfloat[ ]{\includegraphics[width=0.5\textwidth]{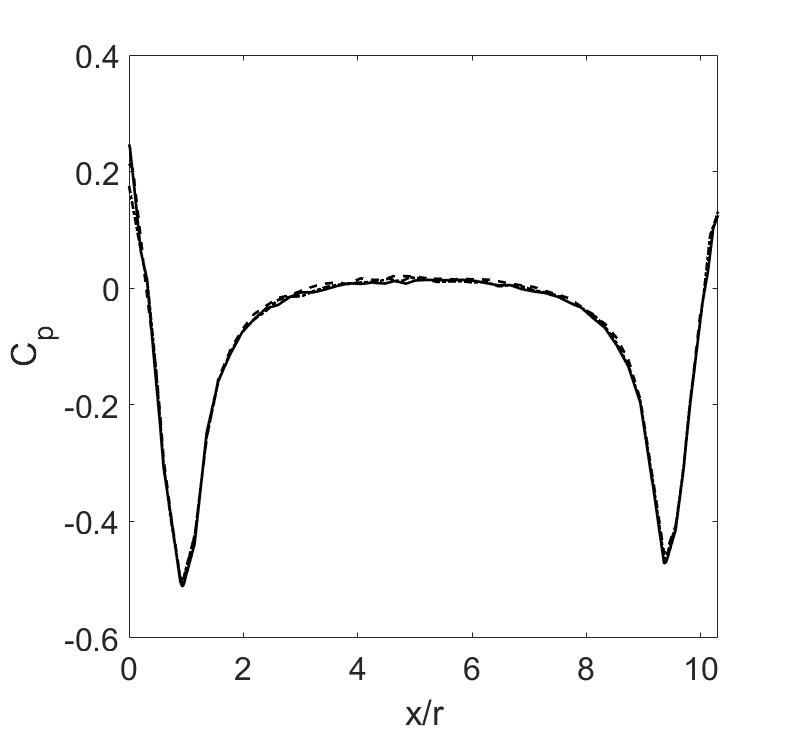}}
 \subfloat[ ]{\includegraphics[width=0.5\textwidth]{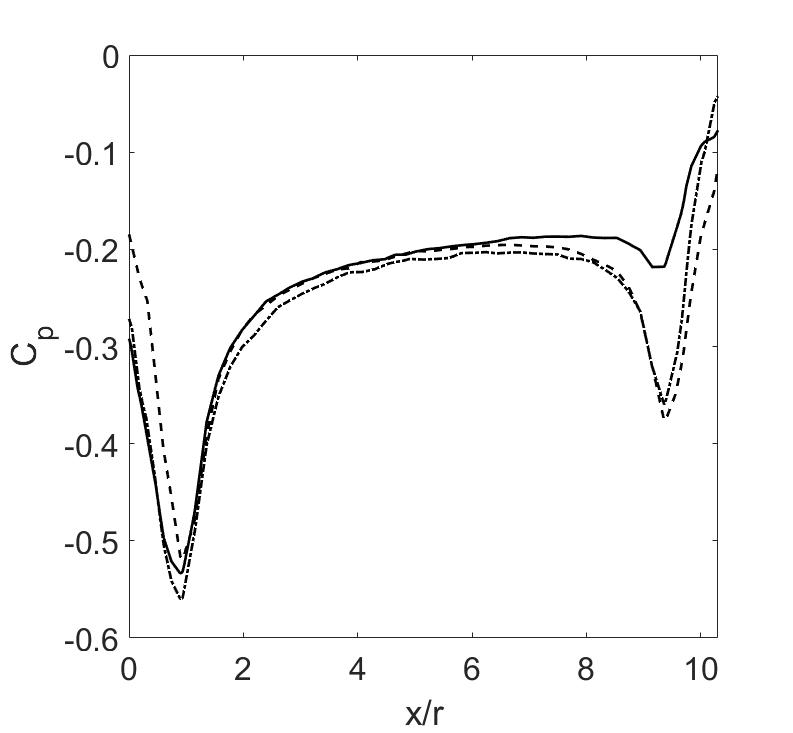}}\\
 \caption{Pressure coefficient evolution on the wall of the AUV along its length away from the grid. Comparison of numerical results for different values of submergence depth. Dashed line $h/r=0$, dashed dot line $h/r=2$ and solid line $h/r=4$. a) Only AUV(OA) case b) Turbulence and AUV (TA) case.\label{fig:13}}
\end{figure}

\begin{figure}
\centering
 \subfloat[ ]{\includegraphics[width=0.5\textwidth]{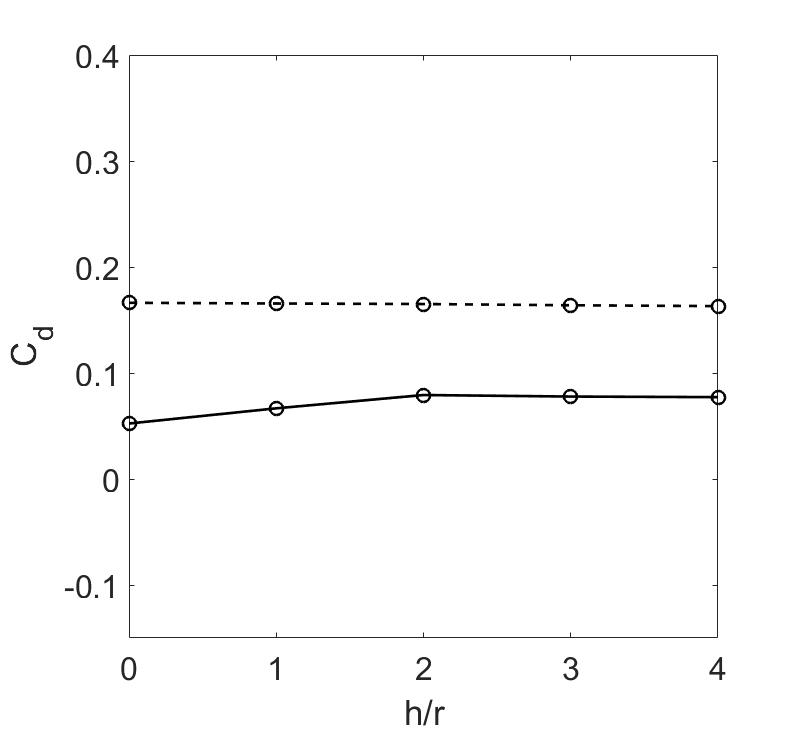}}
 \subfloat[ ]{\includegraphics[width=0.5\textwidth]{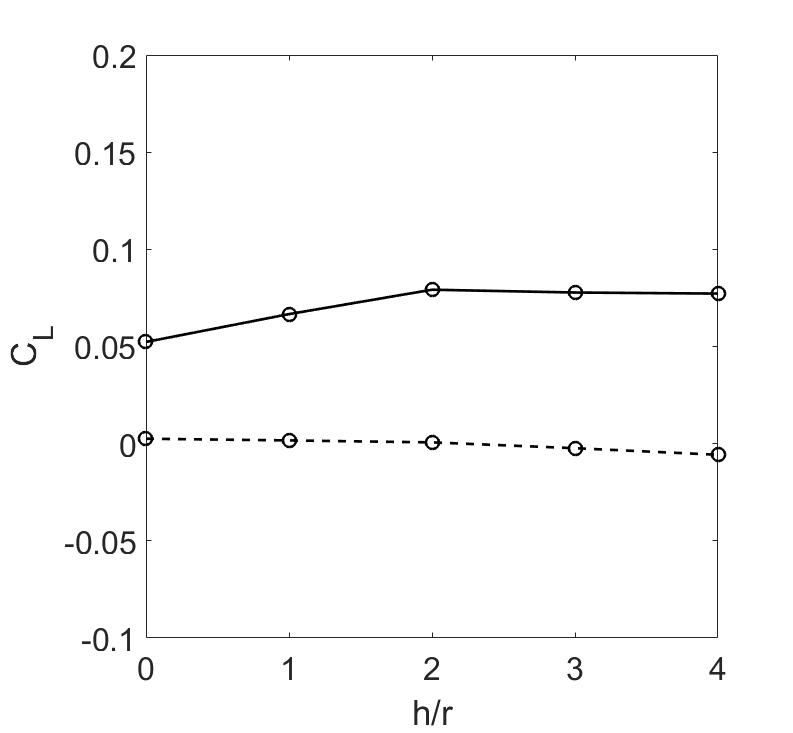}}\\
 \caption{Effect of submergence depth on drag and lift coefficients of the AUV, a) Drag coefficient b) Lift Coefficient. Dashed line OA case and solid line TA case. \label{fig:14}}
\end{figure}

In figure \ref{fig:12} the skin friction coefficient evolution for different submergence depths is presented for both the OA and TA cases. h/r=0 is the central line of the water tank. It is observed that, h/r ratio has a negligible effect on the $C_f$ evolution. However in presence of free stream turbulence, a reduction in $C_f$ value is noticed towards the end of the AUV. The evolution of $C_p$ is shown in figure \ref{fig:13}. It is observed that h/r ratio has negligible effect on the pressure coefficient evolution along the AUV hull. In presence of free stream turbulence at all three h/r ratios the pressure coefficient has a larger magnitude towards the end of the AUV hull. Figure \ref{fig:14} presents the variation of $C_d$ and $C_L$ for different h/r ratio. There is a slight decrease in $C_d$ with increase in h/r ratio. In presence of FST a gradual increase in $C_d$ and $C_L$ is noticed in between h/r=0 and 2.   

\newpage
\section{Concluding remarks}
This study reports results of hydrodynamic coefficients evolution along an AUV hull at three different volumetric Reynolds numbers using high fidelity RSM based simulations. The RSM predictions of turbulent flow field were validated against the experimental results of turbulence kinetic energy and Reynolds stresses obtained from the experiments in the recirculating water tank. From the RSM simulations it is observed that the hydrodynamic coefficients were very responsive to applied turbulence flow fields and with increase in turbulence levels the drag coefficient decreases. A critical comparison of flow evolution along the AUV hull for different submergence depths and angle of attacks were also performed. The  drag and lift coefficients increases with angle of attacks, however slight variation of those coefficients is observed at different submergence depths. The experimental and numerical results presented in this paper can be utilized as a training data set for the design modification of powering and maneuvering system of the AUV operating in oceans and rivers with varied levels of turbulence. The comparison between the experimental data and the CFD simulations suggest that Reynolds Stress Models may be a viable alternative for the design and optimization of AUVs, especially under complex turbulent flows. 





\newpage
\bibliographystyle{elsarticle-num}
\bibliography{asme2e.bib}







\end{document}